\documentclass[prl,amsmath,amssymb,twocolumn]{revtex4}

\usepackage{graphicx}
\usepackage{subfigure}
\usepackage{dcolumn}
\usepackage{verbatim}
\usepackage{epstopdf}

\def\dd{\textrm{d}}
\def\Tr{\textrm}
\def\Bf{\boldsymbol}

\def\rv{\Bf{r}}
\def\kv{\Bf{k}}

\def\nv{\Bf{n}}
\def\Gv{\Bf{G}}
\def\xh{\hat{\Bf{x}}}
\def\yh{\hat{\Bf{y}}}
\def\zh{\hat{\Bf{z}}}

\begin{document}
\title{Superfluid-insulator transitions of the Fermi gas\\ with near-unitary interactions
in a periodic potential}

\author{Eun Gook Moon, Predrag Nikoli\'c,  and Subir Sachdev}
\affiliation{Department of Physics, Harvard University, Cambridge MA
02138}

\begin{abstract}
We consider spin-1/2 fermions of mass $m$ with interactions
near the unitary limit. In an applied periodic
potential of amplitude $V$ and period $a_{\Tr{L}}$, and with a
density of an even integer number of fermions per unit cell, there
is a second-order quantum phase transition between superfluid and
insulating ground states at a critical $V=V_{\Tr{c}}$. We compute the universal ratio
$V_{\Tr{c}} m a_{\Tr{L}}^2/\hbar^2$
 at $N=\infty$ in a model with Sp($2N$) spin
symmetry. The insulator interpolates between a band insulator of
fermions and a Mott insulator of fermion pairs. We discuss implications for recent experiments.
\end{abstract}

\date{\today}

\maketitle

An important milestone in the studies of ultracold atoms has been
the observation of superfluidity in degenerate gases of fermionic
atoms across a two-body Feshbach resonance
\cite{jin0,martin0,chin,martin1,jin}. As a function of the detuning,
$\nu$, across the resonance, these systems interpolate from a
Bose-Einstein condensate of diatomic molecules for large negative
$\nu$, to a Bardeen-Cooper-Schrieffer (BCS) paired state of a Fermi
liquid for large positive $\nu$. For $\nu \approx 0$, neither
limiting description applies, and we have a superfluid state of
fermions with interactions near the unitarity limit. As has been
emphasized in recent work \cite{unitary,rvs,Diehl} this entire crossover
has striking universal aspects, with all physical properties
determined only by $\nu$ and the density of the Fermi gas.

In a separate development, ultracold gases of bosonic atoms were
placed in an optical lattice potential \cite{Greiner}. With
increasing lattice depth, the bosons exhibited a
superfluid-to-insulator quantum phase transition.

In the quest to realize strongly correlated quantum phases of
interacting fermions, a recent experiment \cite{MIToptlat} has
combined the techniques of the above experiment by studying fermions
of mass $m$ with near unitary interactions in the presence of an
optical lattice potential of period $a_{\Tr{L}}$ and amplitude $V$.
This paper shall demonstrate that the universality arguments can be
extended to include the periodic potential after including a single
energy scale associated with $V$. In particular, we find a universal
phase diagram shown in Fig~\ref{phdiag} below. This phase diagram
has superfluid-insulator quantum phase transitions with a density of
an even integer number ($n_{\Tr{F}}$) of fermions per unit cell at
critical amplitude $V=V_{\Tr{c}}$ which obeys
\begin{equation}
V_{\Tr{c}} = \frac{\pi^2\hbar^2}{4ma_{\Tr{L}}^2} F_{n_{\Tr{F}}} ( a_{\Tr{L}} \nu)
\label{eq:vc}
\end{equation}
where $F_{n_{\Tr{F}}}$ is a universal function of $a_{\Tr{L}} \nu$
dependent only on the even integer $n_{\Tr{F}}$. The transition at
unitarity occurs at $F_{n_{\Tr{F}}}(0)$. We will show that
$F_{n_{\Tr{F}}}$ can be determined in a $1/N$ expansion in a model
with Sp($2N$) spin symmetry. Explicit numerical results for the
universal phase diagram and the function $F_{n_{\Tr{F}}}$ will be
presented below at $N= \infty$.

An important aspect of the physics accounted for by our analysis is
that the insulator near the critical point is a novel quantum state
which is neither a band insulator of fermions, nor a Mott insulator
of bosonic fermion pairs. Instead, it is in a interesting
intermediate regime in which multiple single-particle bands are
occupied. Previous computations of ultra-cold atoms in optical
lattices have relied on effective tight-binding models \cite{Duan,
Gubbels, Koetsier, Koponen}, but such approaches are not expected to
be quantitatively accurate near the transition to the superfluid. A
recent computation \cite{ZhaiHo} uses a method related to ours, but does
not account for the off-diagonal couplings between different reciprocal
lattice vectors in Eq.~(\ref{FreeEnergy}) below; these are essential for
a proper result, and make the computation much more demanding.
Our $1/N$ expansion is also able to quantitatively account for the
strong interactions between the fermions in the multiple bands. One
consequence of the many occupied bands is that the computational
requirements are demanding even at the leading $N=\infty$ level. So
our present numerical results will be limited to $N=\infty$ although
we will set up a formalism that allows computations to all orders in
$1/N$. It is also worth noting that previous studies of ground state
properties \cite{rvs} found that $1/N$ corrections were quite small
in the unitarity limit, $\nu \approx 0$.

We consider $2N$ species of fermionic atoms $\psi_{i \sigma}$
($i=1\dots N, \sigma\in\lbrace \uparrow,\downarrow \rbrace$) coupled
to a single field $\Phi$ of s-wave Cooper pairs, or
molecules. This is an Sp($2N$) generalization of the popular
``two-channel'' model, and the physical case $N=1$ can be accessed
in $1/N$ expansions \cite{unitary, rvs}. The atoms experience an
optical lattice potential $V(\rv)$; we choose a simple cubic latice
potential
\begin{displaymath}
V(\rv) = V \left\lbrack \cos^2\left(\frac{\pi
x}{a_{\Tr{L}}}\right) + \cos^2\left(\frac{\pi y}{a_{\Tr{L}}}\right) +
\cos^2\left(\frac{\pi z}{a_{\Tr{L}}}\right)
  \right\rbrack,
\end{displaymath}
and our computations have simple generalization to other lattice
structures. The density of the fermions is controlled by chemical
potential $\mu$ and we assume that spin polarization is zero. The
imaginary-time action of this many-body system, which includes all
terms for a description of the universal physics in the vicinity of
a two-body Feshbach resonance is:
\begin{eqnarray}\label{FullAction}
\mathcal{S} & = & \int \dd\tau \dd^3 r \Bigl\lbrack \psi_{i
\sigma}^{\dagger}
  \left( \frac{\partial}{\partial\tau}
  - \frac{\nabla^2}{2m}-\mu+V(\rv) \right) \psi_{i \sigma}^{\phantom{\dagger}} \\
  & & + \Phi^{\dagger} \psi_{i \downarrow}^{\phantom{\dagger}}
       \psi_{i \uparrow}^{\phantom{\dagger}}
  + \Phi \psi_{i \uparrow}^{\dagger}
       \psi_{i \downarrow}^{\dagger}
\Bigr\rbrack +  N \frac{m\nu}{4\pi} \int \dd\tau \dd^3 r\Phi^\dagger\Phi
\nonumber
\end{eqnarray}
Note that the Cooper pair field $\Phi$ is actually a Hubbard-Stratonovich field that decouples the fermion interaction terms, and hence does not have a bare dispersion, or Berry's phase. Thanks to this, we have
conveniently rescaled the $\Phi$ field to absorb in it the
interaction coupling. Among the relevant operators is
detuning $\nu$ from the Feshbach resonance in the absence of the
lattice, and the co-efficient that contains it is fixed by relating
the scattering matrix of this theory to the scattering length
$a=-1/\nu$ for $V(\rv)=0$.

Our primary result is that the above model has a universal phase
diagram as a function of $\mu/E_{\Tr{r}}$,  $E_{\Tr{r}}/V$, and $\nu a_{\Tr{L}}$,
where $E_{\Tr{r}} = \pi^2\hbar^2 /(4 m a_{\Tr{L}}^2)$ is the molecular recoil energy. The phase boundaries are shown in Fig.~\ref{phdiag} as a function of the first
two parameters for different values of $\nu a_{\Tr{L}}$.

We will explore the superfluid phase boundary by focusing on the
superfluid order parameter. We integrate out the fermion fields
$\psi_{i \sigma}$ and obtain the effective action of the Cooper pair
field $\Phi$, which can be expressed using Feynman diagrams:
\begin{eqnarray}\label{Action}
&& \mathcal{S}_{\Tr{eff}} = N \frac{m\nu}{4\pi} \int \dd\tau \dd^3 r
   \Phi^\dagger\Phi + \\[-0.4cm]
&& ~ \raisebox{-0.3cm}{\includegraphics[height=0.8cm]{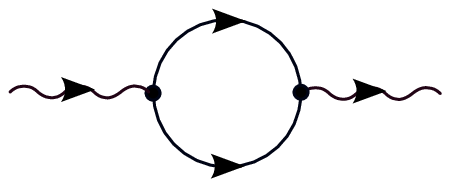}}
+
  \frac{1}{2} \raisebox{-0.63cm}{\includegraphics[height=1.5cm]{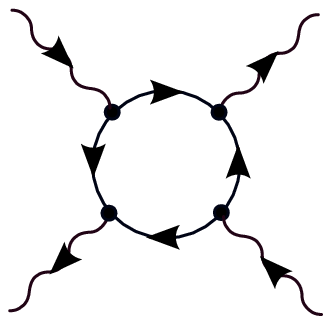}} +
  \frac{1}{3} \raisebox{-0.82cm}{\includegraphics[height=1.8cm]{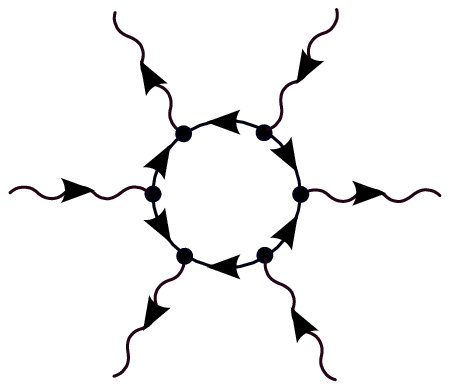}}
  + \cdots \ . \nonumber
\end{eqnarray}
Each fermion loop (straight lines) contributes a factor of $N$, so
that $N$ appears as an overall factor in the effective action. Every
external wavy line represents the pair field $\Phi$ or
$\Phi^{\dagger}$, depending on the direction of arrow with respect
to vertex, and every vertex is associated with a point in space and
time. In the present problem, the optical lattice potential $V(\rv)$
breaks translation symmetry, so that momentum is not a good quantum
number. Instead, according to the Bloch's theorem, quantum numbers
are band index $\nv$ and crystal momentum $\kv$ which takes values
within the first Brillouin zone (BZ). Assuming a Bravais optical
lattice with reciprocal vectors $\Gv$, the bare fermion propagator
is given by:
\begin{eqnarray}
G_0(1,2) & = & \sum_{\nv}T\sum_{\omega}
  \int\limits_{\Tr{BZ}}\frac{\dd^3 k}{(2\pi)^3} \sum_{\Gv_1,\Gv_2}
  \frac{\psi^*_{\nv,\kv;\Gv_1} \psi^{\phantom{*}}_{\nv,\kv;\Gv_2}}
    {-i\omega + \epsilon_{\nv,\kv}} \nonumber \\
& \times & e^{i\left(\kv(\rv_2-\rv_1)-\omega(\tau_2-\tau_1)\right)}
  e^{i(\Gv_2\rv_2-\Gv_1\rv_1)} \ ,
\end{eqnarray}
where the Fourier components of the Bloch wavefunctions
$\psi_{\nv,\kv;\Gv}$ and energies $\epsilon_{\nv,\kv}$ are obtained
from the Fourier transformed Schr\"odinger equation:
\begin{eqnarray}\label{Schrod}
&& \sum_{\Gv'} \left\lbrack \left( \frac{(\kv+\Gv)^2}{2m} - \mu
\right)
   \delta_{\Gv,\Gv'} + V_{\Gv-\Gv'}
   \right\rbrack \psi_{\nv,\kv;\Gv'} = \nonumber \\
&& ~~~~ = \epsilon_{\nv,\kv} \psi_{\nv,\kv;\Gv} \qquad (\forall
\kv,\Gv) \ .
\end{eqnarray}
Solving this set of equations numerically requires imposing a
cut-off energy $\Lambda$ that limits the reciprocal vectors kept in
calculations: $\Gv^2 < 2m\Lambda$. For our $V(\rv)$ above, the only
non-zero Fourier components are $V_0=3V/2$, $V_{\pm 2\pi \xh / a_{\Tr{L}}} = V_{\pm
2\pi \yh / a_{\Tr{L}}} = V_{\pm 2\pi \zh / a_{\Tr{L}}} = V/4$.

The presence of superfluidity can be described by a superfluid order
parameter, which due to the optical lattice may have Fourier
components $\Phi_{\Gv}$ at various reciprocal lattice vectors $\Gv$.
Fluctuations $\delta\Phi(\rv)$ of the Cooper pair field are added to
the order parameter, and the total boson field $\Phi(\rv) =
\sum_{\Gv} \Phi_{\Gv}e^{i\Gv\rv} + \delta\Phi(\rv)$ is represented
by the wavy lines in ~(\ref{Action}). For large $N$ the action
$\mathcal{S}_{\Tr{eff}} \propto N$ becomes large, so that
fluctuations $\delta\Phi$ are suppressed; the mean-field theory
becomes exact in the limit $N\to\infty$. Integrating out
$\delta\Phi$ gives rise to corrections of the order $1/N$ to the
mean-field results. This follows from diagrammatic perturbation
theory performed in the effective action, where the first diagram in
~(\ref{Action}), together with the detuning term, defines a `bare'
propagator of the $\Phi$ fields, while all other diagrams define new
vertices of the $\Phi$ fields. Even though the new vertices are
proportional to $N$, the `bare' propagator is proportional to
$1/N$ and hence yields perturbative expansions of thermodynamic
functions in powers of $1/N$.

As in the bosonic case, we expect a second order
superfluid-insulator transition. Near such a transition, the action
terms quadratic in order parameter determine the state, in analogy
to a simple uniform $\Phi^4$ Landau-Ginzburg theory of bosons.
Neglecting the fluctuations $\delta\Phi(\rv)$ in the $N\to\infty$
limit, the free energy density $\mathcal{F}$ is just the effective
saddle-point action divided by volume $\mathcal{V}$ and $\beta=1/T$:
\begin{equation}\label{FreeEnergy}
\frac{\mathcal{F}}{N} =
\frac{\mathcal{S}_{\Tr{eff}}}{N\beta\mathcal{V}} = \sum_{\Gv,\Gv'}
  K^{(2)}_{\Gv,\Gv'} \Phi^{\dagger}_{\Gv} \Phi^{\phantom{\dagger}}_{\Gv'}
  + \mathcal{O}(\Phi^4) \ .
\end{equation}
The quadratic couplings $K^{(2)}_{\Gv,\Gv'}$ can be represented as a
matrix whose components are indexed by the reciprocal lattice
vectors $\Gv$. If this matrix has only positive eigenvalues, the
free energy ~(\ref{FreeEnergy}) is minimized by
$(\forall\Gv)~\Phi_{\Gv}=0$ indicating an insulating ($T=0$)
or normal ($T>0$) phase. Otherwise, the minimum is obtained at
$(\exists\Gv)~\Phi_{\Gv}\neq0$, and the established phase is
superfluid. The role of $\mathcal{O}(\Phi^4)$ terms near a second
order phase transition is only to stabilize the theory.

A naive derivation of $K^{(2)}_{\Gv,\Gv'}$ from the first Feynman
diagram in ~(\ref{Action}) produces an ultra-violet divergent
expression. This divergence stems from the naive continuum form of
the bare field theory, and must be renormalized away by absorbing it
into finite physically measurable renormalized quantities. One step
toward this goal has already been taken by absorbing any bare
molecule mass (a part of the bare molecule dispersion omitted from
~(\ref{FullAction})) into detuning $\nu$, which is measurable and
fixed in the effective field theory simply by the properties of
scattering matrix. The second step is to remove the remaining
unphysical divergent part by dimensional regularization, which at
large momenta $\kv+\Gv$ is carried out just like in a system without
the optical lattice\cite{unitary}. The regularized expression is:
\begin{eqnarray}\label{K2}
&& K^{(2)}_{\Gv,\Gv'} = \frac{m\nu}{4\pi} \delta_{\Gv,\Gv'} +
  \int\limits_{\Tr{BZ}}\frac{\dd^3 k}{(2\pi)^3} \Biggl\lbrace
  \sum_{\Gv''} \frac{m \delta_{\Gv,\Gv'}}{(\kv+\Gv'')^2} - \nonumber \\
&& \sum_{\nv_1,\nv_2} \sum_{\Gv_1,\Gv_2}
  \left\lbrack 1 - f(\epsilon_{\nv_1,\kv}) - f(\epsilon_{\nv_2,-\kv})
    \right\rbrack \times \\
&& \frac{ \left( \psi^*_{\nv_1,\kv;\Gv_1}
             \psi^*_{\nv_2,-\kv;\Gv'-\Gv_1} \right)
        \left(
    \psi^{\phantom{*}}_{\nv_1,\kv;\Gv_2} \psi^{\phantom{*}}_{\nv_2,-\kv;\Gv-\Gv_2}
        \right)}
    {\epsilon_{\nv_1,\kv} + \epsilon_{\nv_2,-\kv}}
  \Biggr\rbrace \nonumber \ ,
\end{eqnarray}
where $f(x) = (1+e^{x/T})^{-1}$ is the Fermi-Dirac distribution
function.

In the following, we numerically compute the matrix $K^{(2)}$ for
reciprocal lattice vectors $|\Gv| < \sqrt{2m\Lambda'}$, where
$\Lambda'\le\Lambda$, in the limit $N\to\infty$. In practice, it is
sufficient to choose a very small $\Lambda'$ ($|\Gv| \le 2\pi
a_{\Tr{L}}^{-1}$), as long as $\Lambda \gg V$. By mapping dependence of the
smallest eigenvalue $\Gamma^{(2)}$ of the matrix $K^{(2)}$ on chemical
potential $\mu$, lattice amplitude $V$ and detuning $\nu$, we find
the second order phase boundary between superfluid and insulating
phases by the condition
\begin{equation}
\Gamma^{(2)} = 0. \label{g1}
\end{equation}
In Fig.\ref{phdiag} we show a contour
plot of the phase boundary as a function of chemical potential and
inverse lattice amplitude at $T=0$ and $N=\infty$.
This choice of plot axes was made in
order to obtain resemblance to the well-known phase diagram
\cite{fwgf} of a superfluid to Mott insulator transition; while
chemical potential directly controls density, inverse lattice
amplitude is correlated with hopping strength $t$ of an effective
tight-binding model. Indeed, insulating regions are dome-shaped and
correspond to integer fillings of optical lattice sites with bosons
(Cooper pairs). The character of an insulator depends on detuning
$\nu$ from the Feshbach resonance. On the BCS side of the resonance,
$\nu>0$, the insulating domes become larger, and converge toward the
fermion band-gap boundaries as $\nu$ is increased. The system then
behaves as a typical band-insulator (weakly paired fermions are insulating
due to a filled band). On the other hand, in the BEC limit $\nu<0$
the domes become `smaller', resembling a Mott insulator. When
molecules are tightly bound, filling up a fermion band is not
sufficient to destroy superfluidity, but repulsion between molecules
needs to step in. Note, however, that true Mott insulating phases, with
an arbitrary integer lattice filling by molecules, cannot be found without
including $1/N$ corrections.

\begin{figure}
\includegraphics[width=2.9in]{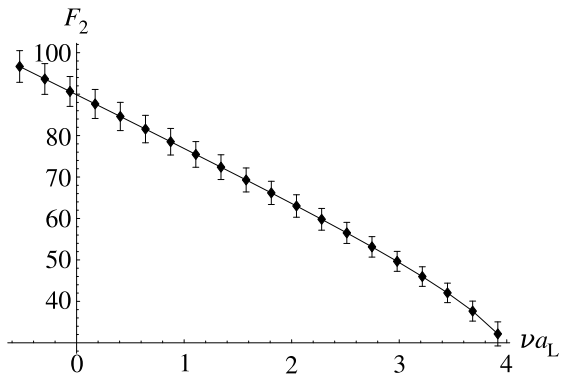}
\caption{\label{qcp}Universal function $F_{n_{\Tr{F}}} ( \nu a_{\Tr{L}} )$
in ~(\ref{eq:vc}) at the superfluid-insulator transition for the first fermion band completely filled, $n_{\Tr{F}}=2$ ($T=0$, $N=\infty$).}
\end{figure}

It is also important to consider the phase diagram as a function of particle
density, rather than chemical potential. The Mott insulating lobes
in Fig.~\ref{phdiag} all have a density of an even integer number,
$n_{\Tr{F}}$, of fermions per unit cell. We therefore fix the
density at $n_{\Tr{F}}$ and study the transition from the insulator
to the superfluid. By generalizing the argument made for the bosonic
case, this $T=0$ transition occurs at the point where we satisfy the
condition (\ref{g1}) along with
\begin{equation}
\frac{\partial \Gamma^{(2)}}{\partial \mu} = 0. \label{g2}
\end{equation}
The two conditions (\ref{g1}) and (\ref{g2}) determine an isolated
point in the phase diagram of Fig.~\ref{phdiag} for each
$n_{\Tr{F}}$, and the location of these points then immediately
yields Eq.~(\ref{eq:vc}). In Fig.~\ref{qcp} we plot the universal function
$F_{n_{\Tr{F}}} ( \nu a_{\Tr{L}})$ for $n_{\Tr{F}}=2$ and $N=\infty$
(calculations were performed with large cut-off $\Lambda\le 10\times (2\pi a_{\Tr{L}}^{-1}))$.

\begin{figure}[!]
\includegraphics[width=3in]{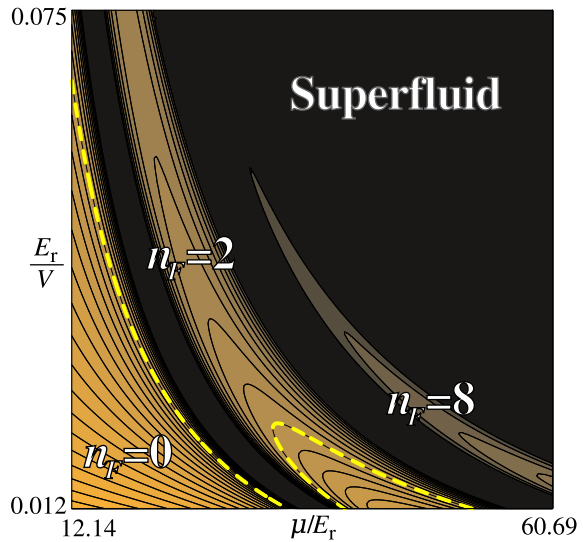}
\caption{\label{phdiag}A contour plot of the superfluid-insulator phase boundaries at $T=0$ and $N=\infty$. Contours, dependent on detuning $\nu$ from the Feshbach resonance, separate superfluid regions surrounding the fermion bands (black) from insulating regions in the band structure gaps. The dashed yellow contours correspond to the transition at the resonance $\nu=0$, while the spacing between contours is $\Delta\nu=a_{\Tr{L}}^{-1}$, where $a_{\Tr{L}}$ is the optical lattice spacing; contours move upward and toward the band edges as $\nu$ grows. Insulators are labeled by the closest even integer to the average filling of the optical lattice by atoms in the grand-canonical ensemble. The reference energy scale is molecule recoil energy $E_{\Tr{r}} = \pi^2\hbar^2/(4m a_{\Tr{L}}^2)$. Accuracy is smaller than that in Fig.\ref{qcp}, because calculating at many points in reasonable time required a small cut-off $\Lambda=3\times(2\pi a_{\Tr{L}}^{-1})$.}
\end{figure}

\begin{figure}[!]
\includegraphics[width=3in]{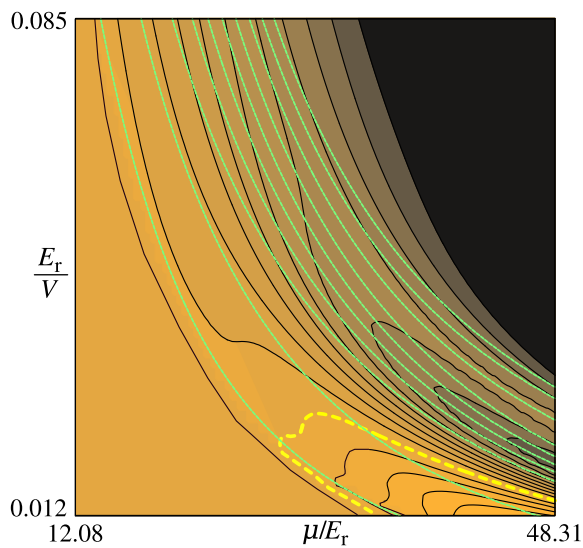}
\caption{\label{phdiag-finiteT}A contour plot of the superfluid-normal phase boundary at $T\approx 1.8 E_{\Tr{r}}$ and $N=\infty$. This temperature is $T\approx 0.6 E_{\Tr{f}}$, where $E_{\Tr{f}}=\hbar^2(6\pi a_{\Tr{L}}^3)^{2/3}/(2m)$ is the Fermi energy of a free fermion gas at the same density as the fermion gas in the lattice with two particles per site. The black and yellow contours are contours of constant detuning, as in Fig~\ref{phdiag}. The green lines are normal-phase constant density lines, starting at $n_{\Tr{F}}=1$ at the left-bottom, displaced by $\Delta n_{\Tr{F}}=1$. The superconductor-normal phase transition can occur at any
density or detuning, at the intersection of the corresponding contour lines.
Note that due to a small cut-off used in plotting this diagram, one should not directly compare it with Fig.~\ref{qcp} - the purpose is only to illustrate the shape of phase boundaries.}
\end{figure}


The obtained values of $V_{\Tr{c}}=F_{n_{\Tr{F}}}(a_{\Tr{L}} \nu) E_{\Tr{r}}$ are larger than those reported in the experiment, $V_{\Tr{c}}\approx 6E_{\Tr{r}}$ \cite{MIToptlat}. However, unlike our computation so far, the experiment is at a non-zero temperature (not known very accurately in the presence of the optical lattice), and thermal fluctuations should
decrease the value of $V_c$. We extended
our results to a range of $T>0$ as shown in Fig.~\ref{phdiag-finiteT}.
The character of the transition at which superconductivity is lost changes qualitatively
at $T>0$. At $T=0$, the insulator must have a density $n_F=\mbox{even integer}$, and so a superconductor-insulator transition can only occur at such values of $n_F$; this was rationale behind the additional constraint in Eq.~(\ref{g2}). However, at the $T>0$, the transition is more properly a superconductor-normal transition, and the normal state can occur at any density.
In Fig~\ref{phdiag-finiteT} we show a contour plot of the boundaries between superfluid and normal phases at finite temperatures ($N=\infty$). As temperature is increased, the non-superfluid domes gradually expand. The contours corresponding to larger values of $\nu$ (BCS-limit) are more affected by thermal fluctuations than those corresponding to smaller values of $\nu$ (BEC-limit). Since at $T>0$ the normal regions can occur when the chemical potential is not in a band gap, the normal regions corresponding to different average lattice fillings can merge when the $\nu$-dependent effect of fluctuations is large enough; when this happens at a particular $\nu$, the appropriate contour stretches all the way from the bottom to the top of the diagram, instead of being dome-shaped. We expect that including molecule fluctuations would increase these effects even further. Now, if an experiment is performed at a fixed density (green lines in Fig.~\ref{phdiag-finiteT}), the phase transition to a normal phase at unitarity (dashed line) can occur at a smaller lattice depth $V_{\Tr{c}}$ than at $T=0$. The reduction of $V_{\Tr{c}}$ can be particularly dramatic if the transition is observed in a region where the $T=0$ non-superfluid regions have merged due to thermal fluctuations. This can even occur at temperatures small compared to the Fermi energy. We suspect this effect is the primary reason for the discrepancy between the $T>0$ experiments and our $T=0$ results.

We thank S.~Diehl for useful discussions, and J.~K.~Chin for helping
us understand better the experiment \cite{MIToptlat}. This research
was supported by the NSF under grant DMR-0537077. E.~G.~Moon is also
supported in part by the Samsung Scholarship.


\begin{thebibliography}{99}

\bibitem{jin0} C.~A.~Regal, M.~Greiner, and D.~S.~Jin,
Phys. Rev. Lett. {\bf 92}, 040403 (2004).

\bibitem{martin0} M.~W.~Zwierlein, C.~A.~Stan, C.~H.~Schunck, S.~M.~F.~Raupach,
A.~J.~Kerman, and W.~Ketterle,
Phys. Rev. Lett. {\bf 92}, 120403 (2004).

\bibitem{chin}
C.~Chin, M.~Bartenstein, A.~Altmeyer, S.~Riedl, S.~Jochim,
J.~H.~Denschlag, and R.~Grimm,
Science {\bf 305} 1128 (2004).

\bibitem{martin1} M.~W.~Zwierlein, J.~R.~Abo-Shaeer, A.~Schirotzek, C.~H.~Schunck,
and W.~Ketterle,
Nature {\bf 435}, 1047 (2005).

\bibitem{jin} J.~Stewart, J.~P.~Gaebler, C.~A.~Regal, and D.~S.~Jin,
Phys. Rev. Lett. {\bf 97}, 220406 (2006).

\bibitem{unitary}
P.~Nikoli\'c and S.~Sachdev,
Phys. Rev. A {\bf 75}, 033608 (2007).

\bibitem{rvs} M.~Y.~Veillette, D.~E.~Sheehy, and L.~Radzihovsky,
Phys. Rev. A {\bf 75}, 043614 (2007).

\bibitem{Diehl}
S.~Diehl and C.~Wetterich, Phys. Rev. A {\bf 73}, 033615 (2006); S.~Diehl, H.~Gies, J.~M.~Pawlowski, C.~Wetterich, cond-mat/0703366 (2007)

\bibitem{Greiner}
M.~Greiner, O.~Mandel, T.~Esslinger, T.~W.~H\"{a}nsch, I.~Bloch,
Nature {\bf 415}, 39 (2002).

\bibitem{MIToptlat}
J.~K.~Chin, D.~E.~Miller, Y.~Liu, C.~Stan, W.~Setiawan, C.~Sanner,
K.~Xu, W.~Ketterle, Nature {\bf 443}, 961 (2006).



\bibitem{Duan}
L.-M.~Duan,
Phys.~Rev.~Lett. {\bf 95}, 243202 (2005).

\bibitem{Gubbels}
K.~B.~Gubbels, D.~B.~M.~Dickerscheid, and H.~T.~C.~Stoof,
New J.~Phys. {\bf 8}, 151 (2006).

\bibitem{Koetsier}
A.~O.~Koetsier, D.~B.~M.~Dickerscheid, and H.~T.~C.~Stoof,
Phys.~Rev.~A {\bf 74}, 033621 (2006).

\bibitem{Koponen}
T.~K.~Koponen, T.~Paananen, J.-P.~Martikainen, and P.~Torma,
cond-mat/0701484.

\bibitem{ZhaiHo}
H.~Zhai, T.-L.~Ho,
cond-mat/0704.2957 (2007).

\bibitem{fwgf} M.~P.~A.~Fisher,
P.~B.~Weichmann, G.~Grinstein, and D.~S.~Fisher,
Phys.~Rev.~B {\bf 40}, 546 (1989).


\end{thebibliography}
\end{document}